\documentclass[a4paper,11pt]{article}
\pdfoutput=1 

\usepackage{jheppub} 

\usepackage[T1]{fontenc} 

\usepackage{slashed}

\allowdisplaybreaks

\subheader{DFPD-2015/TH/27} 
\title{ 
Constrained superfields in Supergravity 
}

\author{Gianguido~Dall'Agata and Fotis~Farakos}

\affiliation{ Dipartimento di Fisica ed Astronomia ``Galileo Galilei''\\
		Universit\`a di Padova, Via Marzolo 8, 35131 Padova, Italy}
		
\affiliation{INFN, Sezione di Padova \\
		Via Marzolo 8, 35131 Padova, Italy}
  
\abstract{
We analyze constrained superfields in supergravity. 
We investigate the consistency and solve all known constraints, presenting a new class that may have interesting applications in the construction of inflationary models.
We provide 
the  superspace Lagrangians for minimal supergravity models based on them 
and write the corresponding theories in component form using a simplifying gauge for the goldstino couplings.  
}

\begin{document} 
\maketitle
\flushbottom

\section{Introduction} 
\label{sec:introduction}

Supersymmetry breaking is at the core of any realistic physical scenario that makes use of superstring or supergravity models.
Unfortunately, we do not have a full understanding of all possible supersymmetry breaking mechanisms yet and, when this happens at a high energy scale, one may question the validity of approaches using supergravity to address particle physics or cosmology.
However, non-linear realizations are a very useful tool to explore consequences of spontaneously broken symmetries within an effective Lagrangian approach.
They become especially appealing when the mass splitting between fields in the same multiplet is large compared to the typical energy scale of the model one is analyzing.
In the superfield language this is realized by imposing consistent constraints on the superfields, so that some of their components are going to be expressed as functions of the others.
It is therefore of clear interest to understand and study the dynamics of such constrained superfields both in global and local supersymmetry.

While in the case of global supersymmetry there is a good understanding of many different constraints \cite{Rocek:1978nb,Ivanov:1978mx,Ivanov:1982bpa,Casalbuoni:1988xh,Komargodski:2009rz,Antoniadis:2010hs,
Kuzenko:2011ti,Kuzenko:2011tj,Dudas:2011kt,Antoniadis:2011xi,Dudas:2012fa,Farakos:2015vba}, 
there is still no general analysis available for constrained superfields in supergravity theories \cite{Lindstrom:1979kq,Samuel:1982uh,Farakos:2013ih,Dall'Agata:2014oka,Dall'Agata:2015zla,Antoniadis:2014oya,
Dudas:2015eha,Ferrara:2014kva,Ferrara:2015gta,Hasegawa:2015bza,
Antoniadis:2015ala,Kuzenko:2015yxa,
Bergshoeff:2015tra,Kallosh:2014hxa,
Hasegawa:2015era,Kallosh:2015tea,Kallosh:2015pho,Kallosh:2014via,Kallosh:2015sea}. 
With this work we aim at filling this gap, by analyzing known consistent globally supersymmetric constraints within the supergravity framework, providing their solutions and discussing the corresponding supergravity actions in superfield formalism.
These constraints are non-dynamical and they are imposed directly on the superfields, so that they maintain linear supersymmetry transformations between the composite fields and the fundamental ones, which means that they can be directly combined with standard superspace methods. 
On the other hand,  supersymmetry is realized non-linearly  on the fundamental component fields, 
once their superpartners are removed from the spectrum. 

Constrained superfields in supergravity have been recently the subject of intense scrutiny because of their possible applications to cosmology \cite{AlvarezGaume:2010rt,AlvarezGaume:2011xv,Schillo:2015ssx,Antoniadis:2014oya,Ferrara:2014kva,Carrasco:2015iij,Kallosh:2014hxa,Kallosh:2014via,Dall'Agata:2014oka,Ferrara:2015cwa,Kahn:2015mla} and to brane supersymmetry breaking scenarios \cite{Bandos:2015xnf,Aparicio:2015psl,Kallosh:2015nia,Bergshoeff:2015jxa}.
For this reason, we will also discuss the supergravity actions that include such constrained superfields using the explicit component language, though we will do it by using a simplifying gauge choice that removes a certain component of the goldstino field $(G_{\alpha} = 0)$.
We will see that in such gauge, the expressions for the Lagrangian in the minimal matter coupled setup become quite simple and make the task of discussing physical applications easy.
It is in principle not clear that all supersymmetric constraints should have consistent solutions in a supergravity setup. 
For example some of them are not algebraic (in the sence that super-covariant derivatives are involved in the inversion procedure), or supergravity auxiliary fields may also appear.  
Therefore the importance in  solving these constraints is that one proves their consistency by explicit calculation. 

As will become clear in the following, the solution to the constraints in supergravity can become much more complicated than in the corresponding globally supersymmetric case. 
In fact, interesting differences emerge for some of the constraints, and this can play an important role in physical applications.
We therefore analyze the constraints in a somewhat increasing order of complexity, comparing the new supergravity solutions to the corresponding ones in the rigid limit.
We will start by reviewing the simplest constrained nilpotent superfield \cite{Rocek:1978nb,Lindstrom:1979kq,Casalbuoni:1988xh,Komargodski:2009rz}
\begin{equation}
X^2 = 0, 
\end{equation}
which is used to remove the superpartner of the goldstino field (the sGoldstino \cite{Brignole:1998uu}) in the supersymmetry breaking sector.
We then consider combinations of constrained superfields.
First we will complete the analysis of \cite{Dall'Agata:2015zla}, by discussing models with two chiral constrained superfields \cite{Brignole:1997pe}:
\begin{equation}
	X^2 = 0, \qquad X Y =0.
\end{equation}
These models have the goldstino superfield coupled to another chiral field where the scalar field component has also been removed.
We will then discuss models where the second constrained chiral superfield is a U(1) gauge superfield with field-strength $W_{\alpha}$.
The gaugino is consistently removed from the spectrum by imposing 
\begin{equation}
X {\cal W}_{\alpha} = 0.
\end{equation}
We then discuss models with constrained superfields where the fermionic degrees of freedom have been removed.
The first possibility is to consider the coupling between the nilpotent goldstino superfield $X$ and a second chiral superfield ${\cal H}$ and require that the combination $X \overline{\cal H}$ be chiral:
\begin{equation}\label{XH}
\overline {\cal D}_{\dot \alpha} \left(X\overline {\cal H}\right) = 0.
\end{equation}
This constraint has also the effect of removing the auxiliary field in ${\cal H}$. 
In addition, one may also impose a more stringent constraint that removes also one real scalar degree of freedom by combining the second chiral superfield ${\cal A}$  with $X$ so that $X \overline{\cal A}$ is chiral and at the same time identical to $X {\cal A}$:
\begin{equation}
\label{XAintr}
X \, {\cal A} = X \, \overline {\cal A}.
\end{equation}
This constraint  leaves only a real propagating mode in the ${\cal A}$ multiplet, removing the other scalar, the fermion and the auxiliary field in terms of the goldstino.

Finally, we construct a new constrained chiral superfield ${\cal C}$, where only its fermion is a function of the goldstino, leaving both the scalar and the auxiliary field non trivial.
This is obtained by imposing the constraint
\begin{equation}
X \overline X \, {\cal D}_\alpha {\cal C} = 0,
\end{equation}
where $X$ is the constrained goldstino superfield.
This is interesting because, differently from the other models that remove fermions, the supergravity scalar potential can be obtained by setting to zero the sgoldstino scalar in the standard formula, while in the other models its form is not the standard one.

All calculations have been done using the standard 2-component superfield language \cite{Wess:1992cp} and following the standard procedure for building a supergravity action outlined in the same book. 

\medskip

\textbf{Note added:} While this work was under completion, \cite{Ferrara:2015tyn,Carrasco:2015iij} appeared in the ArXiv, with a discussion of cosmological models with constrained superfields satisfying (\ref{XH})--(\ref{XAintr}). This partially overlaps the content of our section \ref{sec:models_constraining_fermions}.


\section{Scalarless models} 
\label{sec:scalarless_models}

\subsection{Sgoldstinoless models} 
\label{sub:sgoldstinoless}

A chiral superfield in supergravity satisfies the covariant constraint condition
\begin{equation}
	\overline{\cal D}_{\dot \alpha} X = 0.
\end{equation}
When expanded in terms of the supergravity $\Theta$ variables \cite{Wess:1992cp}, one obtains the covariant components:
\begin{equation}
X = x + \sqrt 2\, \Theta^\alpha G_\alpha + \Theta^2 F^x . 
\end{equation}
When supersymmetry gets broken, in the absence of vector multiplets, one can assume that the goldstino $G_{\alpha}$ is the fermion in such multiplet and that the scalar $x$ is its superpartner, the sgoldstino, while $F^x$ gives the order parameter of supersymmetry breaking.
In models where the sgoldstino is not present in the effective theory, we can still use supergravity to describe the low-energy action by constraining the superfield $X$ with a quadratic algebraic condition:
\begin{equation}\label{X2}
	X^2 = 0.
\end{equation}
An efficient method to obtain the constraints imposed by such condition in supergravity is to solve the equations that arise by hitting it with the highest possible number of covariant chiral derivatives and extracting the $\theta = \overline \theta= 0$ projection.
In this case
\begin{equation}
	{\cal D}^2 (X^2)|_{\theta = \overline \theta = 0} = 2[X {\cal D}^2 X + {\cal D}^\alpha X {\cal D}_{\alpha} X]|_{\theta = \overline \theta = 0} = -8 x F^x + 4 G^\alpha G_{\alpha} = 0,
\end{equation}
which is easily solved whenever $F^x \neq 0$ by setting $x = \frac{G^2}{2 F^x}$.
The other constraints follow from ${\cal D}_{\alpha} (X^2)|_{\theta = \overline \theta = 0}$ and from $X^2|_{\theta = \overline \theta = 0}$\footnote{From now on  ${\cal F}|$ will stand for ${\cal F}|_{\theta = \overline \theta = 0}$, where ${\cal F}$ is a generic superfield.}, but they are simply consistency conditions for the solution we just found. 
This implies that the solution to the nilpotency constraint (\ref{X2}) is identical in form to the one of global supersymmetry:
\begin{equation}
X = \frac{G^2}{2 F^x} + \sqrt 2\, \Theta^\alpha G_\alpha + \Theta^2 F^x . 
\end{equation}

The Lagrangian describing the supergravity couplings of such multiplet in superspace is (we set $M_P = 1$ for the time being)
\begin{equation}
\label{L1}
{\cal L} = \int d^2 \Theta \, 2 {\cal E}\, \left[ \frac38 (\overline {\cal D}^2 - 8 {\cal R}) \text{e}^{-K/3} + W \right] + c.c.,
\end{equation}
where the K\"ahler potential and superpotential have forms that are constrained by (\ref{X2}) to
\begin{equation}\label{KahlX}
K = X \overline X 
\end{equation}
and 
\begin{equation}\label{WforX}
W = f X + W_0,  
\end{equation}
respectively, for arbitrary complex parameters $f$ and $W_0$. 
One should then carefully integrate out the auxiliary field, which does not appear anymore in a simple quadratic expression\footnote{Note that we could have redefined the fermion component of the $X$ superfield with an arbitrary function of the auxiliary field. 
The constraint can still be consistently solved and the action resulting by integrating over superspace coordinates can have once again the auxiliary field appearing only quadratically. 
In this basis, however, the ``rescaled'' goldstino kinetic term is not canonically normalized and proper rescalings by the expectation value of the auxiliary field brings the action back to the form computed in \cite{Bergshoeff:2015tra,Hasegawa:2015bza}.} due to the constraint, to get the final form of the Lagrangian in components.
The final form has been computed with different methods in \cite{Bergshoeff:2015tra,Hasegawa:2015bza}.
Here we simply note that, by using the fact that the goldstino is a pure gauge degree of freedom of supersymmetry
\begin{equation}
\delta G_\alpha = \sqrt{2} \,\epsilon_\alpha f + \cdots,
\end{equation}
we may always perform a gauge fixing so that we remove it from the final action:
\begin{equation}
\label{G=0}
G_\alpha = 0 . 
\end{equation}
In this gauge the component Lagrangian simplifies drastically to
\begin{equation}
\begin{split}
e^{-1} {\cal L} = &  -\frac12 R\left(e_m^a,\psi_m\right) 
+ \frac{1}{2} \epsilon^{klmn} (\overline \psi_k \overline \sigma_l {\cal D}_m \psi_n - \psi_k \sigma_l {\cal D}_m \psi_n) 
\\[2mm]
&  
- \overline{W_0} \, \overline \psi_a \overline \sigma^{ab} \overline \psi_b 
- W_0 \, \psi_a  \sigma^{ab}  \psi_b 
- |f|^2 + 3 |W_0|^2  
\end{split}
\end{equation}
and supersymmetry is spontaneously broken on a vacuum that can be Minkowski, de Sitter or anti-de Sitter according to the value of the $f$ and $W_0$ constants.

Note also that coupling gravity to matter fields promotes all derivatives appearing in the action to super-covariant ones.
Since their expression is going to be useful in the following, we give here such derivatives for the constrained Goldstino sector:
\begin{equation}
\widehat D_m \left( \frac{G^2}{2 F} \right) =  \partial_m \left( \frac{G^2}{2 F} \right) 
- \frac{1}{\sqrt 2} \psi_m^\beta G_\beta,
\end{equation}
\begin{equation}
\widehat D_m G_\alpha =  {\cal D}_m G_\alpha 
- \frac{1}{\sqrt 2} \psi_{m \alpha} F 
- \frac{i}{\sqrt 2} \overline \psi_m^{\ \dot \beta} \sigma^c_{\alpha \dot \beta} \widehat D_c \left( \frac{G^2}{2 F} \right)\,. 
\end{equation}


\subsection{Two constrained chiral superfields: $X^2 =0 = XY$} 
\label{sub:two_constrained_chiral_superfields_x_2_0_xy}

If, in addition to the nilpotent goldstino multiplet, there is another chiral superfield
\begin{equation}
Y = y + \sqrt 2 \Theta^\alpha \chi_\alpha + \Theta^2 F^y  
\end{equation}
whose scalar component gets a sufficiently high mass to remove it from the low-energy spectrum, we can describe the resulting model by constraining $Y$,  setting \cite{Brignole:1997pe,Komargodski:2009rz}
\begin{equation}
\label{XY}
X \, Y = 0 . 
\end{equation}
Also this constraint is algebraic and once more, the action of two chiral covariant spinorial derivatives gives a constraint that fixes the scalar $y$ in terms of the other fields:
\begin{equation}
	{\cal D}^2 (XY)| = -4 F^x y - 4 x F^y + 4 G^\alpha \chi_{\alpha} = 0.
\end{equation}
This proves that also in this case the solution remains the same as in global supersymmetry \cite{Komargodski:2009rz,Dall'Agata:2015zla}:
\begin{equation}
Y = \frac{G^\alpha \chi_\alpha}{F} - \frac{G^2}{2 F^2} F^y   
+ \sqrt 2 \Theta^\alpha \chi_\alpha + \Theta^2 F^y . 
\end{equation}
In fact hitting the constraint with a lower number of derivatives gives only consistency conditions, which are identically satisfied.

A general model of $X$ and $Y$ superfields coupled to supergravity is given by a 
theory with Kaehler potential \cite{Dall'Agata:2015zla}
\begin{equation}
\label{KXY}
K = X \overline X + Y \overline Y + a (X \overline Y^2 + \overline X Y^2) + b (Y \overline Y^2 + \overline Y Y^2) + c Y^2 \overline Y^2  
\end{equation}
and superpotential
\begin{equation}
\label{WXY}
W = W_0 + f X + g Y + h Y^2 . 
\end{equation}
We note that in these models the goldstino is not simply the fermion in $X$, due to the term proportional to $g$ in the superpotential. 
As noted in \cite{Dall'Agata:2015zla}, the physical goldstino is a linear combination of $G$ and $\chi$, proportional to 
\begin{equation}
f G + g \chi . 
\end{equation}
Since this field is a pure gauge degree of freedom in our model, we can always fix it to any value that is useful to simplify our calculations.
It is easy to realize that the solutions to the superfield constraints that we impose create composite fields that are always dependent on combinations of the fermion field in the goldstino superfield $X$.
This means that setting $G_{\alpha} = 0$ provides a simplifying gauge choice where most of these composite fields are going to be set to zero.
For instance, in the current case, this gauge choice simplifies the solution to the constraints so that the constrained superfields $X$ and $Y$ reduce to 
\begin{equation}
X|_{G=0} = \Theta^2 F^x, \quad  Y|_{G=0} = \sqrt 2 \Theta^\alpha \chi_\alpha + \Theta^2 F^y . 
\end{equation}
This gauge choice has the drawback that one still has residual goldstino components in the action and one should therefore remove them from the fermion mass matrix to obtain the physical masses, but overall the advantage of having to deal with a simpler action from the beginning guides us to this choice.

Once this gauge has been imposed, the Lagrangian for the component fields becomes
\begin{equation}
\label{lfg}
\begin{split}
e^{-1} {\cal L} = & - \frac12 R + \epsilon^{klmn} \overline \psi_k \overline \sigma_l {\cal D}_m \psi_n 
- i \overline \chi \overline \sigma^m {\cal D}_m \chi  
\\[2mm]
&
+ \frac14 \left[ i \epsilon^{klmn} \psi_k \sigma_l \overline \psi_m 
+ \psi_m \sigma^n \overline \psi^m \right] \chi \sigma_n \overline \chi 
\\[2mm]
&
- \frac18 \left[ 1 - 8 (c - a^2 - b^2) \right] \chi^2 \overline \chi^2 
- \frac{i}{\sqrt 2} g \, \chi \sigma^a \overline \psi_a 
- \frac{i}{\sqrt 2} g \, \overline \chi \overline \sigma^a  \psi_a 
\\[2mm]
& - W_0 ( \psi_a \sigma^{ab} \psi_b + \overline \psi_a \overline \sigma^{ab} \overline \psi_b ) 
- (f^2 + g^2 - 3 W_0^2) 
\\[2mm]
& - (h - af - bg) \,  \chi^2 - (h - af - bg) \,  \overline \chi^2 . 
\end{split}
\end{equation}
The scalar potential in this model is constant and vanishes when  
\begin{equation}
\label{Wfg}
3 W_0^2 = f^2 + g^2. 
\end{equation}
In this case we get a Minkowski vacuum and we can explicitly check the results in \cite{Dall'Agata:2015zla} for the mass term of the physical fermion as well as compute for the first time its quartic coupling.
In order to do so, we have to disentangle the physical fermion from the gravitino in the mass matrix, which is done by performing the shift 
\begin{equation}
\label{shift}
\psi_{m \alpha} \rightarrow \psi_{m \alpha} - \frac{i g}{3 \sqrt 2 W_0} \sigma_{m \alpha \dot \alpha}  \overline \chi^{\dot \alpha} . 
\end{equation}
This shift will eliminate the mass cross terms between the gravitino and the fermion, but on the other hand it will create new contributions to the various couplings, including the kinetic term for $\chi$. 
Note that, since part of the Goldstino has already been gauged away, this shift is not the conventional one (see for example \cite{Wess:1992cp}).
Due to our gauge choice, we also need to perform a shift of the form $\delta_\text{shift} \psi_{m \alpha} \sim \partial_m \chi_\alpha $ that removes non-diagonal kinetic terms between the gravitino and $\chi_{\alpha}$.  
After these operations, we can then compute the physical mass term, which is
\begin{equation}
e^{-1} {\cal L} \supset  - \frac{1}{f^2} \left[ (f^2 + g^2)(h - af  - bg) - g^2 m_{3/2} \right] (\chi^2 + \overline \chi^2),
\end{equation}
while the gravitino mass is $m_{3/2} = W_0 = \sqrt{(f^2 + g^2) / 3}$. 
We can also explicitly compute the physical 4-fermi interaction, which is  
\begin{equation}
e^{-1} {\cal L} \supset  \frac{g^2 + f^2}{f^4}   
\left\{  
\frac{g^2}{12} \left(  \frac{g^2}{2 (f^2 + g^2)} - 1 \right) 
- \frac{f^2 + g^2}{8} \left[ 1 - 8 (c - a^2 - b^2) \right] 
\right\} \chi^2 \overline \chi^2 . 
\end{equation}
Note that when we set $g=0$ the Goldstino direction is completely aligned with $G_\alpha$, and the gauge choice $G_\alpha = 0$ leads directly to diagonal mass terms and canonical normalized kinetic terms as can be seen from \eqref{lfg}. 


\subsection{Models with a single U(1) vector} 
\label{sub:models_with_a_single_u_1_vector}

Another interesting model that contains no scalar fields is given by a supergravity theory where we couple the goldstino superfield to a U(1) vector multiplet, described by the real superfield $V$ and its covariant field-strength
\begin{equation}
{\cal W}_\alpha = -\frac{1}{4} ( \overline {\cal D}^2 - 8 {\cal R}) {\cal D}_\alpha V . 
\end{equation}
The standard kinetic term for such multiplet is 
\begin{equation}
\label{Lg}
{\cal L}_{\cal W} = \frac{1}{4 g^2} \int d^2 \Theta \, 2 {\cal E} \, {\cal W}^2 + c.c.,
\end{equation}
which gives a model with a massless gaugino and a massless U(1) vector. 

When supersymmetry breaking leads to a very high mass for the gaugino, we can remove if from the low-energy spectrum by imposing the constraint 
\begin{equation}
\label{XW}
X \, {\cal W}_\alpha = 0 . 
\end{equation}

As we did before, we can read the constraint on the component fields by acting with the square of the chiral covariant derivative.
We obtain the expression\footnote{Note that we are assuming here that we \emph{first} solve the constraint $X^2 = 0$ and \emph{then} the constraint $X {\cal W}_{\alpha} =0$. In fact there are other possible consistent solutions (in the global case), where $X$ is given as a function of $W^2$. These solutions preserve linear $N=1$ supersymmetry and describe the goldstino multiplet of an additional broken supersymmetry \cite{Bagger:1994vj,Rocek:1997hi,Antoniadis:2008uk}.}
\begin{equation}
\label{lambda}
\begin{split}
\lambda_\alpha = & - \frac{i}{2} \widetilde G^2 \left[ \sigma^{c}_{\alpha \dot \beta} \widehat D_c \overline \lambda^{\dot \beta} 
-\frac{i}{2} (\lambda_\alpha \overline M + b_{\alpha}^{\dot \beta} \overline \lambda_{\dot \beta}) \right] 
\\
& + \frac{i}{\sqrt{2}} \widetilde G^\beta \left[ -2i \sigma^{ba \ \gamma}_{\ \ \alpha} \epsilon_{\gamma \beta} \widehat D_b v_a 
+ \epsilon_{\alpha \beta}  \text{D}  \right]  ,
\end{split}
\end{equation}
where
\begin{equation}
\widetilde G_\alpha = \frac{G_\alpha}{F}.
\end{equation}
There are two noteworthy things happening here.
First, the constraint contains the gaugino field on both sides of the equation and one must therefore solve it by an iterative procedure, replacing all gaugini appearing in the solution by the right hand side of (\ref{lambda}), until the expressions that contain them vanish because of the nilpotency of the $G$ fields.
Second and most important, the auxiliary fields of the gravity multiplet (namely $M$ and $b$) enter for the first time explicitly in the relation determining $\lambda$.

The solution to \eqref{lambda} is
\begin{align}
\nonumber
\lambda_\sigma = &  \frac{1}{4} \sigma^a_{\sigma \dot \phi} \epsilon^{\dot \phi \dot \sigma} \sigma^b_{\tau \dot \sigma} \epsilon^{\tau \alpha} I_{ab\alpha}
- \frac{i}{2} \widetilde G^2 \sigma^a_{\sigma \dot \phi} \epsilon^{\dot \phi \dot \sigma} e_a^m {\cal D}_m\left( \overline A_{\dot \sigma} \overline {\widetilde G}^2 \right)
\\[2mm] 
\nonumber
& 
- \frac{i}{2} \sigma^a_{\sigma \dot \phi} \epsilon^{\dot \phi \dot \sigma} \overline B^{\ \alpha}_{\dot \sigma} \widetilde G^2 \overline {\widetilde G}^2 
\Gamma_{\alpha}^{\ \tau} e_{a}^{m} {\cal D}_m \widetilde G_\tau 
- \frac{i}{2} \widetilde G^2 \sigma^a_{\sigma \dot \phi} \epsilon^{\dot \phi \dot \sigma} 
e_{a}^{m} {\cal D}_m(\overline {\widetilde G}_{\dot \tau} \overline \Gamma_{\dot \sigma}^{\ \dot \tau} ) 
\\[2mm] 
\nonumber
& -\frac{i}{2} \widetilde G^2 \sigma^a_{\sigma \dot \phi} \epsilon^{\dot \phi \dot \sigma} 
\overline {\widetilde G}_{\dot \tau} \overline \Delta_{\dot \sigma}^{\ \dot \tau \alpha} 
\Gamma_{\alpha}^{\ \tau} e_{a}^{m} {\cal D}_m \widetilde G_\tau 
+ \frac{i}{2} \widetilde G^2 \sigma^a_{\sigma \dot \phi} \epsilon^{\dot \phi \dot \sigma} 
\overline {\widetilde G}_{\dot \tau} \overline \Delta_{\dot \sigma}^{\ \dot \tau \alpha} 
\Delta_{\alpha}^{\ \tau \dot \epsilon}  \overline {\widetilde G}_{\dot \rho} \overline \Gamma_{\dot \epsilon}^{\ \dot \rho} 
e_{a}^{m} {\cal D}_m \widetilde G_\tau 
\\[2mm] 
\nonumber
& + \frac{i}{2} \widetilde G^2 \sigma^a_{\sigma \dot \phi} \epsilon^{\dot \phi \dot \sigma} 
e_{a}^{m} {\cal D}_m (\overline Z_{\dot \tau \dot \alpha} \overline E_{\dot \sigma}^{\ \dot \tau \dot \alpha}) 
+ \frac{i}{2} \widetilde G^2 \overline {\widetilde G}^2 B_{\sigma}^{\ \dot \sigma}  \sigma^a_{\alpha \dot \sigma} \epsilon^{\alpha \gamma} 
\Gamma_{\gamma}^{\ \omega}  e_{a}^{m} {\cal D}_m \widetilde G_\omega  
\\[2mm] 
& + A_\sigma \widetilde G^2 
+ \widetilde G^2 \overline {\widetilde G}^2 B_{\sigma}^{\ \dot \sigma} \overline A_{\dot \sigma} 
+ \widetilde G^2 B_{\sigma}^{\ \dot \sigma} \overline \Gamma_{\dot \sigma}^{\ \dot \tau} \overline {\widetilde G}_{\dot \tau}  
- \widetilde G^2 B_{\sigma}^{\ \dot \sigma} \overline E_{\dot \sigma}^{\ \dot \tau \dot \epsilon} \overline Z_{\dot \tau \dot \epsilon}    
\\[2mm] 
\nonumber
& -\frac{i}{2} \widetilde G_\tau \Delta_{\sigma}^{\ \tau \dot \epsilon} \overline {\widetilde G}^2  
\sigma^a_{\phi \dot \epsilon} \epsilon^{\phi \alpha}  
 e_{a}^{m} {\cal D}_m( \Gamma_{\alpha}^{\ \gamma} \widetilde G_\gamma  ) 
+ \frac{i}{2} \widetilde G_\tau \Delta_\sigma^{\ \tau \dot \epsilon} 
\overline {\widetilde G}^2  \sigma^a_{\phi \dot \epsilon} \epsilon^{\phi \alpha} 
\widetilde G_\gamma \Delta_{\alpha}^{\ \gamma \dot \rho} \overline \Gamma_{\dot \rho}^{\ \dot \gamma} 
e_{a}^{m} {\cal D}_m \overline {\widetilde G}_{\dot \gamma}   
\\[2mm] 
\nonumber
& + \Gamma_{\sigma}^{\ \tau} \widetilde G_\tau  
- \widetilde G_\tau \Delta_{\sigma}^{\ \tau \dot \epsilon} \overline A_{\dot \epsilon} \overline {\widetilde G}^2   
- \Delta_{\sigma}^{\ \tau \dot \epsilon} \overline {\widetilde G}^2 Z_{\tau \alpha} \overline B_{\dot \epsilon}^{\ \alpha} 
- \widetilde G_\tau \Delta_{\sigma}^{\ \tau \dot \epsilon} 
\overline {\widetilde G}_{\dot \tau} \overline \Gamma_{\dot \epsilon}^{\ \dot \tau}   
\\[2mm] 
& - \Delta_{\sigma}^{\ \tau \dot \epsilon} \overline {\widetilde G}_{\dot \tau} Z_{\tau \alpha} 
\overline \Delta_{\dot \epsilon}^{\ \dot \tau \alpha}  
+  \widetilde G_\tau \Delta_{\sigma}^{\ \tau \dot \epsilon} \overline Z_{\dot \tau \dot \alpha} 
E_{\dot \epsilon}^{\ \dot \tau \dot \alpha} 
- \widetilde G_\tau E_{\sigma}^{\ \tau \epsilon} \widetilde G_\phi  \Gamma_{\epsilon}^{\ \phi}  
+ E_{\sigma}^{\ \tau \epsilon} H_{\tau \phi \dot \alpha} \Delta_{\epsilon}^{\ \phi \dot \alpha} , \nonumber
\end{align}
where Goldstino-dependent terms have been collected in 
\begin{equation}
\begin{split}
I_{cd \sigma} =  {\widetilde G}^2 e_{c}^{m} {\cal D}_m \left( \overline {\widetilde G}^2 e_{d}^{n} {\cal D}_n \left[ 
\phantom{\frac12} \!\!\!\!\!\! \right.  \right.
& 
- \frac{i}{2}  {\widetilde G}^2 \sigma^{a}_{\sigma \dot \phi} \epsilon^{\dot \phi \dot \sigma} \overline \Gamma_{\dot \sigma}^{\ \dot \tau} 
e_{a}^{k} {\cal D}_k \overline {\widetilde G}_{\dot \tau} 
+A_\sigma {\widetilde G}^2 
+\Gamma_\sigma^{\ \tau} {\widetilde G}_\tau  
\\[2mm]
& 
\left. \left. \phantom{\frac12} \!\!\!\!\! 
- {\widetilde G}_\tau \Delta_{\sigma}^{\ \tau \dot \epsilon} \overline {\widetilde G}_{\dot \tau} \overline \Gamma_{\dot \epsilon}^{\ \dot \tau} 
- {\widetilde G}_\tau E_{\sigma}^{\ \tau \epsilon} {\widetilde G}_\phi \Gamma_{\epsilon}^{\ \phi} \right] \right) 
\end{split}
\end{equation}
and 
\begin{align}
H_{\tau \phi \dot \alpha} =&  {\widetilde G}_\tau {\widetilde G}_\phi \left( 
\frac{i}{2} \overline {\widetilde G}^2 \sigma^c_{\beta \dot \alpha}\, \epsilon^{\beta \alpha} 
\Gamma_\alpha^{\ \gamma} e_{c}^{m} {\cal D}_m {\widetilde G}_\gamma 
+ \overline A_{\dot \alpha} \overline {\widetilde G}^2 
+ \overline \Gamma_{\dot \alpha}^{\ \dot \gamma} \overline {\widetilde G}_{\dot \gamma} 
+ \overline {\widetilde G}_{\dot \gamma} \overline E_{\dot \alpha}^{\ \dot \gamma \dot \rho} 
\overline {\widetilde G}_{\dot \sigma} \overline \Gamma_{\dot \rho}^{\ \dot \sigma} 
\right) ,
\\
\overline Z_{\dot \tau \dot \alpha} = &  \overline {\widetilde G}_{\dot \tau} \overline {\widetilde G}_{\dot \gamma} 
\overline \Gamma_{\dot \alpha}^{\ \dot \gamma} 
-\frac{i}{2} \overline {\widetilde G}_{\dot \tau} \overline {\widetilde G}_{\dot \gamma} 
\overline \Delta_{\dot \alpha}^{\ \dot \gamma \rho} 
{\widetilde G}^2 \sigma^c_{\rho \dot \beta} \epsilon^{\dot \beta \dot \omega} \overline \Gamma_{\dot \omega}^{\ \dot \epsilon} 
e_{c}^{m} {\cal D}_m \overline {\widetilde G}_{\dot \epsilon}  
+ \overline {\widetilde G}_{\dot \tau} \overline {\widetilde G}_{\dot \gamma} \overline\Delta_{\dot \alpha}^{\ \dot \gamma \rho}  
A_\rho {\widetilde G}^2 
\\[2mm]
& + \overline {\widetilde G}_{\dot \tau} \overline {\widetilde G}_{\dot \gamma} \overline \Delta_{\dot \alpha}^{\ \dot \gamma \rho} 
{\widetilde G}_\gamma \Gamma_{\rho}^{\ \gamma} 
+ \overline {\widetilde G}_{\dot \tau} \overline {\widetilde G}_{\dot \gamma} \overline \Delta_{\dot \alpha}^{\ \dot \gamma \rho} 
{\widetilde G}_\gamma \Gamma_{\epsilon}^{\ \omega} 
{\widetilde G}_\omega \overline E_{\rho}^{\ \gamma \epsilon}  , \nonumber
\end{align}
while the Goldstino-independent terms have been collected in
\begin{align}
A_\alpha = & \frac14 \sigma^c_{\alpha \dot \beta} \overline \psi_c^{\ \dot \beta} \text{D} 
+ \frac{i}{2}  \sigma^c_{\alpha \dot \beta} \overline \sigma^{db \, \dot \beta}_{\ \ \ \dot \kappa} \overline \psi_c^{\ \dot \kappa} 
\left\{ e_d^m {\cal D}_m v_b + \frac{i}{2} \psi_d \slashed{v} \overline \psi_b \right\} ,
\\
B_{\alpha}^{\ \dot \alpha} =& 
\frac{1}{4} \sigma^c_{\alpha \dot \beta} 
\overline \sigma^{db \, \dot \beta}_{\ \ \ \dot \kappa} \overline \psi_c^{\ \dot \kappa} 
\overline \sigma_{b}^{\dot \alpha \gamma} \psi_{d \gamma} - \frac14 b_{\alpha}^{\, \dot \alpha} ,
\\
\Gamma_{\alpha}^{\ \gamma} =& \sqrt{2} \sigma^{ba \ \gamma}_{\ \ \alpha} 
\left\{ e_b^m {\cal D}_m v_a + \frac{i}{2} \psi_b v \overline \psi_a \right\} 
+ \frac{i}{\sqrt{2}} \text{D} \delta_{\alpha}^{\gamma} ,
\\
\Delta _{\alpha}^{\ \gamma \dot \rho} =& \frac{i}{\sqrt{2}} \sigma^{ba \ \gamma}_{\ \ \alpha} 
\overline \sigma_{a}^{\dot \rho \rho} \psi_{b \rho} ,
\\
E_{\alpha}^{\ \gamma \rho} =& \frac{i}{\sqrt{2}} \sigma^{ba \ \gamma}_{\ \ \alpha} 
\overline \psi_{b \dot \rho}  \overline \sigma_{a}^{\dot \rho \rho} . 
\end{align}
Finally, the supercovariant derivatives appearing in the previous formulae are
\begin{equation}
\begin{split}
\widehat D_m \overline \lambda^{\dot \beta} &= {\cal D}_m \overline \lambda^{\dot \beta} 
+\frac{i}{2} \overline \psi_m^{\dot \beta} \text{D} 
- \overline \sigma^{db \, \dot \beta}_{\ \ \dot \kappa} \overline \psi_m^{\dot \kappa} \widehat D_d v_b ,
\\
\widehat D_m v_{\alpha \dot \alpha}&= 
 {\cal D}_m v_{\alpha \dot \alpha} 
+ i (\psi _{m \alpha} \overline \lambda_{\dot \alpha} + \overline \psi_{m \dot \alpha} \lambda_\alpha ) 
+\frac{i}{2} \psi_m \slashed{v} \overline \psi_a \sigma^a_{\alpha \dot \alpha} 
. 
\end{split}
\end{equation}

While the solution in the locally supersymmetric case is much more complicated than the one obtained in global supersymmetry, we checked that by setting $b_m=0$, $M=0$, $\psi_m^{\ \alpha} =0$ and $e_{m}^{a}=\delta_{m}^{a}$ we recover the result presented in \cite{Komargodski:2009rz}. 
This completes the direct proof that the constraint $X{\cal W}_{\alpha} =0$ is consistent also in supergravity.

When writing down the general action for this sector, we should note that due to the constraint \eqref{XW}, the most general coupling of the form 
\begin{equation}
\int d^2 \Theta \, 2 {\cal E} \, H(X)  {\cal W}^2 + c.c. 
\end{equation}
will always reduce to \eqref{Lg} (if it does not vanish). 
This means that the coupling of this constrained U(1) multiplet to supergravity is going to be described by a Lagrangian of the form
\begin{equation}
{\cal L}_X + {\cal L}_{\cal W},
\end{equation}
where ${\cal L}_X$ is the Lagrangian in (\ref{L1}) for the K\"ahler potential and superpotential in (\ref{KahlX}) and (\ref{WforX}).
Once again  the gauge choice $G_\alpha=0$ simplifies dramatically the terms appearing in the Lagrangian.
In this gauge
\begin{equation}
{\cal W}_{\alpha} |_{G=0} = \Theta_{\alpha} D - i  (\sigma^{ab}\Theta)_{\alpha} \widehat{D}_a v_b + \Theta^2 \sigma^a_{\alpha \dot \beta}\left(\frac{i}{2} \overline \psi_a^{\dot \beta} \text{D} 
- \overline \sigma^{db \, \dot \beta}_{\ \ \dot \kappa} \overline \psi_a^{\dot \kappa} \widehat D_d v_b\right).  
\end{equation}
The form of the Lagrangian in components is then
\begin{equation}
\begin{split}
e^{-1} {\cal L} =&  -\frac12 R 
+ \frac{1}{2} \epsilon^{klmn} (\overline \psi_k \overline \sigma_l {\cal D}_m \psi_n - \psi_k \sigma_l {\cal D}_m \psi_n) 
- \frac{1}{4 g^2} F^{mn} F_{mn} 
\\
& - W_0 \, \overline \psi_a \overline \sigma^{ab} \overline \psi_b 
- W_0 \, \psi_a  \sigma^{ab}  \psi_b 
- f^2 + 3 W_0^2,
\end{split}
\end{equation}
where $F_{mn} = \partial_m v_n - \partial_n v_m$.



\section{Models constraining fermions} 
\label{sec:models_constraining_fermions}

The constraints we presented in the previous section were such that all scalars in the models were removed from the spectrum and in one case also the gaugino.
In the following we will consider other constraints that leave some scalars in the spectrum, while removing fermion fields.

Originally these constraints were introduced having in mind phenomenological models for physics beyond the Standard Model \cite{Komargodski:2009rz}, where one would like to have at least one light scalar, the Higgs field, in the spectrum of the low-energy effective theory.
Clearly scalar fields could also be of direct interest for cosmology, when one deals with inflationary scenarios.
In fact the second constraint we will discuss, while proposed in a different context \cite{Komargodski:2009rz}, has also been used for inflation \cite{Kahn:2015mla,Ferrara:2015tyn,Carrasco:2015iij}.

In this section we will prove that these constraints have consistent solutions in supergravity, where once more the composite fields contain new terms and couplings due to gravitational interactions, but coincide with the known expressions in the rigid limit.

\subsection{Surviving complex scalar} 
\label{sub:surviving_complex_scalar}

The first model we analyze contains the goldstino multiplet $X$ coupled to a second chiral superfield, whose covariant components are
\begin{equation}
{\cal H} = H + \sqrt 2 \,\Theta^\alpha \psi^H_\alpha + \Theta^2 F^H.
\end{equation}
To remove the fermion in ${\cal H}$ one could couple $X$ to the chiral derivative of ${\cal H}$, so that $\psi_{\alpha}^H$ becomes a function of the goldstino.
This is expressed by the request that $X \overline{\cal H}$ be a chiral superfield:
\begin{equation}\label{dXH}
	\overline{\cal D}_{\dot \alpha} \left(X \overline{\cal H}\right) = 0.
\end{equation}
This constraint is not chiral and therefore the conditions on the components can be obtained by applying the maximum number of chiral and anti-chiral covariant derivatives that do not annihilate identically the left hand side of (\ref{dXH}).
The projection $\overline{\cal D}^{\dot \alpha} {\cal D}^2[\overline{\cal D}_{\dot \alpha} \left(X \overline{\cal H}\right) ]| = 0$ gives an equation that fixes the auxiliary field $F^H$ in terms of the other fields of the multiplets.
After an iterative procedure like the one applied in previous sections we obtain a closed expression for $F^H$, which is
\begin{equation}
\label{FH}
\overline F^H = \overline F^{H(0)} - i \frac{G \sigma^c \overline \psi_c }{2 \sqrt 2 F^x} \overline F^{H(0)}  
\end{equation}
where 
\begin{equation}
\begin{split}
\overline F^{H(0)} = & \frac{1}{16 F^x} \Big{[} \,  \overline \psi_H^{\dot \alpha} \, (8 i  \sigma^c_{\gamma \dot \alpha} \widehat D_c  G^\gamma 
- 8 b_{\gamma \dot \alpha} G^\gamma) 
+ 16  \widehat D^c \overline H \, \widehat D_c \left( \frac{G^2}{2 F^x} \right) 
\\
& \ \ \ \  \ \ \   
+ 8 i G^\alpha \sigma^m_{\alpha \dot \rho} \epsilon^{\dot \rho \dot \alpha} 
\left\{ {\cal D}_m \overline \psi^H_{\dot \alpha} + \frac{i}{\sqrt 2} \psi_m^\rho \sigma^c_{\rho \dot \alpha} \widehat D_c \overline H   \right\} \Big{]} 
\end{split}
\end{equation}
and
\begin{equation} 
\widehat D_m H =  \partial_m H - \frac{1}{\sqrt 2} \psi_m^\beta \psi^H_\beta. 
\end{equation}

The projection	${\cal D}^2 \left[ X  \overline {\cal D}_{\dot \alpha}  \overline {\cal H} \right]|= 0$ produces an equation for $\psi_a^H$ in terms of $H$, which, after one iteration to invert the relation gives
\begin{equation}
\label{hH}
\overline \psi^H_{\dot \alpha} = - i \frac{G^\alpha}{F^x} \sigma^m_{\alpha \dot \alpha} {\cal D}_m \overline H 
-  \frac{G^\alpha}{\sqrt 2 F^x} \sigma^a_{\alpha \dot \alpha} \overline \psi_a^{\dot \beta} 
\frac{G^\rho}{F^x}  \sigma^n_{\rho \dot \beta} {\cal D}_n \overline H .
\end{equation}
Note that in this last expression the goldstino appears always in the combination $G^\alpha / F^x $. 
All the other projections are then identically satisfied.

These solutions differ from the globally supersymmetric case because of the appearance of the supergravity auxiliary field $b$ and of various terms involving the gravitino.
However, we checked that these results coincide with those in \cite{Komargodski:2009rz} in the rigid limit.

If we want to discuss the supergravity action that one obtains by using these constrained superfields we can start from the most general K\"ahler potential ans superpotential, which are
\begin{equation}
	K = |X|^2 + X \, P({\cal H, \overline{\cal H}}) + \overline X \, P({\cal H, \overline{\cal H}}) + Z({\cal H, \overline{\cal H}})
\end{equation}
and
\begin{equation}
	W = g({\cal H}) + X\, f({\cal H}).
\end{equation}
Note that any term of the form $|X|^2 {\cal F}({\cal H}, \overline{\cal H})$ in the K\"ahler potential can be reabsorbed by a field redefinition $X^\prime = X \sqrt{\cal F}$, because the constraint (\ref{XH}) implies that $X^\prime$ is also chiral and satisfies $(X^\prime)^2 = 0 = X^\prime \overline{\cal D}_{\dot \alpha} \overline{\cal H} = 0$.
This redefinition may generate a new contribution in the superpotential of the form $ X^\prime \tilde f({\cal H}, \overline{\cal H})$, which is also chiral thanks to the constraint on ${\cal H}$, but it can be reabsorbed in the second and third terms of the K\"ahler potential by a K\"ahler transformation.

Also in this instance the gauge choice $G_{\alpha} = 0$ simplifies the constrained superfield expansion to
\begin{equation}
	X|_{G = 0} = \Theta^2 F^x, \qquad {\cal H}|_{G = 0} = H,
\end{equation}
so that $X$ is effectively the goldstino multiplet and contains the order parameter of supersymmetry breaking.
Using this expansion in the general superspace action we obtain the Lagrangian in components
\begin{equation}
	\begin{split}
	e^{-1} {\cal L} = & -\frac12 R 
	+ \frac{1}{2} \epsilon^{klmn} (\overline \psi_k \overline \sigma_l {\cal D}_m \psi_n - \psi_k \sigma_l {\cal D}_m \psi_n) 
	-Z_{{\cal H} \overline {\cal H}} \, \partial^m H \, \partial_m \overline{H}
	- V 
	\\[2mm]
	& + \frac14 \epsilon^{klmn} Z_{{\cal H}} \, \psi_l \sigma_m \overline \psi_n \, \partial_k H
	 + \frac14 \epsilon^{klmn} Z_{\overline{\cal H}} \, \psi_l \sigma_m \overline \psi_n \, \partial_k \overline{H} 
	- \text{e}^{Z/2} (g \, \overline \psi_a \overline \sigma^{ab} \overline \psi_b + \overline g \, \psi_a \sigma^{ab} \psi_b ),
	\end{split}
\end{equation}
where
\begin{equation}
	V = e^Z\left(|f+ g P|^2 - 3 |g|^2\right).
\end{equation}
Note that the vanishing of the $F^H$ auxiliary field in this gauge implies that the scalar potential cannot be simply obtained from the standard component action by setting to zero the sgoldstino field.
One should therefore be careful when extending this result to models with more matter multiplets.
 

\subsection{Real scalar model} 
\label{sub:real_scalar_model}

It has been suggested \cite{Komargodski:2009rz,Kahn:2015mla} that, further constraining the ${\cal H}$ superfield just discussed   one could remove also one of the two real scalars in the lowest component of the constrained chiral superfield.
This could be of interest when considering systems with approximate or exact shift symmetries, or in cosmological models where we want to identify the residual scalar with the inflaton.

Following \cite{Komargodski:2009rz} we therefore introduce a chiral superfield
\begin{equation}
{\cal A} = a + i \Sigma + \sqrt 2 \Theta^\alpha \psi^A_\alpha + \Theta^2 F^{\cal A}
\end{equation}
and couple it to the goldstino superfield $X$, requiring that $X \overline{\cal A}$ be chiral and identical to $X {\cal A}$:
\begin{equation}
\label{XA}
X {\cal A} - X \overline {\cal A} = 0 . 
\end{equation}
As in the previous section the projection with ${\cal D}^2 \overline{\cal D}^2$ constrains $F^{\cal A}$ in terms of the other fields of the two multiplets as in (\ref{FH}) and the projection with ${\cal D}^2 \overline{\cal D}$ constrains $\psi^A_{\alpha}$ in the same way as (\ref{hH}).
The novelty comes from the ${\cal D}^2$ projection
\begin{equation}
	{\cal D}^2 X ({\cal A} - \overline{\cal A})| + 2 {\cal D}^\alpha X {\cal D}_{\alpha} {\cal A}| + X {\cal D}^2 {\cal A}| = 0,
\end{equation}
which gives
\begin{equation}
\Sigma = i \frac{ G^2}{2 (F^x)^2} F^{\cal A} 
- \frac{i}{2 F^x}  G^{ \alpha}  \psi^A_{\alpha}, 
\end{equation}
which also implies
\begin{equation}
\Sigma = - i \frac{\overline G^2}{2 (\overline F^x)^2} \overline F^{\cal A} 
+ \frac{i}{2 \overline F^x} \overline G_{\dot \alpha} \overline \psi_A^{\dot \alpha} . 
\end{equation}
Once we replace the expressions for $F^{\cal A}$ and $\psi^A_{\alpha}$ we obtain
\begin{equation}
	\label{bGa}
	\begin{split}
	\Sigma = & \  \text{Re} V^c {\cal D}_c a - \text{Im} L^{ab} {\cal D}_a {\cal D}_b \left[ \text{Re} V^c {\cal D}_c \right] 
	\\
	& - \text{Im} V^a {\cal D}_a \left[ \text{Re} V^c {\cal D}_c a  
	- \text{Im} V^c {\cal D}_c \left( \text{Re}  V^e {\cal D}_e a \right) \right],
	\end{split}
\end{equation}
where 
\begin{equation}
\begin{split}
\overline V^a = & - i \frac{\overline G^2}{64 (F^x)^2 (\overline F^x)^2} \left(1 - i \frac{G \sigma^b \overline \psi_b}{2 \sqrt 2 F^x} \right) 
\Big{[} -8 (i \widehat D_c G^\beta \sigma^c_{\beta \dot \beta} - G^\beta \sigma^c_{\beta \dot \beta} b_c) 
- 8 \sqrt 2 \widehat D_c \left( \frac{G^2}{2 F^x} \right) \overline \psi_{\ \dot \beta}^{c}  
\\[2mm]
& 
\ \ \ \ \ \ \ \ \ \ \ \  \ \ \ \ \ \ \ \ \ \ \ \  \ \ \ \ \ \ \ \ \ \ \ \  \  \ 
-4(G \sigma^e \overline \sigma^c \psi_e) \overline \psi_{c \dot \beta} \Big{]} 
\left( i \overline \sigma^{a \dot \beta \rho} G_\rho 
+ \frac{G^2}{2 \sqrt 2 F^x} \overline \sigma^{f \dot \beta \rho} \sigma^a_{\rho \dot \rho} \overline \psi_f^{\ \dot \rho}  \right) 
\\[2mm]
& 
-i \frac{\overline G^2}{16 F^x (\overline F^x)^2} \left(1 - i \frac{G \sigma^b \overline \psi_b}{2 \sqrt 2 F^x} \right)  
\Big{[} 4  \widehat D^a\left( \frac{G^2}{2 F^x} \right) 
+ \sqrt 2 G \sigma^e \overline \sigma^a \overline \psi_e   \Big{]} 
\\[2mm]
&
+ \frac{i}{2 \overline F^x} \overline G_{\dot \beta}  \left( 
\frac{i}{F^x} \overline \sigma^{a \dot \beta \rho} G_\rho + \frac{G^2}{2 \sqrt 2 (F^x)^2} 
\overline \sigma^{c \dot \beta \rho} \sigma^a_{\rho \dot \rho} \overline \psi_c^{\ \dot \rho}  
\right) 
\\[2mm]
& 
+ \frac{\overline G^2}{8 F^x (\overline F^x)^2}  \left(1 - i \frac{G \sigma^b \overline \psi_b}{2 \sqrt 2 F^x} \right)  
G^\alpha \sigma^b_{\alpha \dot \beta} {\cal D}_b\left( 
\frac{i}{F^x} \overline \sigma^{a \dot \beta \rho} G_\rho 
+ \frac{G^2}{2 \sqrt 2 (F^x)^2} \overline \sigma^{e \dot \beta \rho} \sigma^a_{\rho \dot \rho} \overline \psi_e^{\ \dot \rho} 
\right) 
\end{split}
\end{equation}
and
\begin{equation}
\overline L^{ab} = - i \frac{G^2 \overline G^2}{8 (F^x)^2 (\overline F^x)^2} \eta^{ab}. 
\end{equation}
We checked that also in this case the rigid limit reproduces the expression given in \cite{Komargodski:2009rz}.

The most general coupling of ${\cal A}$ with the nilpotent superfield $X$ in supergravity is described by a K\"ahler potential of the form
\begin{equation}
\label{KPAX}
K = X \overline X + Z({\cal A}, \overline {\cal A})
\end{equation}
and superpotential
\begin{equation}
	W = g({\cal A}) +  X f({\cal A}). 
\end{equation}
In fact also in this case we can canonically normalize the $|X|^2$ term by a field redefinition on ${\cal A}$.
Moreover, the constraint (\ref{XA}) tells us that we can replace any $\overline{\cal A}$ by ${\cal A}$ in any function that is mulitplied by $X$ and viceversa for any function that is multiplied by $\overline X$.
This means that we can move any term linear in $X$ into the K\"ahler potential to the superpotential, where it redefines the $f({\cal A})$ function\footnote{As noted in \cite{Ferrara:2015tyn}, ${\cal Z}$ can also be restricted to ${\cal Z} = h({\cal A} + \overline{\cal A}) ({\cal A} - \overline{\cal A})^2$.
This follows by noting that $({\cal A} - \overline{\cal A})$ satisfies a nilpotency constraint of degree 3 and removing the first two coefficients in the expansion of the K\"ahler potential by a customary K\"ahler transformation.}. 

Also in this case supersymmetry is broken when $F^x \neq 0$ and we can use the $G_{\alpha} =0$ gauge, such that 
\begin{equation}
X|_{G=0} = \,  \Theta^2 F^x \qquad {\cal A}|_{G=0} = \, a 
\end{equation}
and obtain the general Lagrangian in component form:
\begin{equation}
\begin{split}
e^{-1} {\cal L} = & -\frac12 R 
+ \frac{1}{2} \epsilon^{klmn} (\overline \psi_k \overline \sigma_l {\cal D}_m \psi_n - \psi_k \sigma_l {\cal D}_m \psi_n) 
-Z_{{\cal A} \overline {\cal A}} \, \partial^m a \, \partial_m a 
- V 
\\[2mm]
& + \frac14 \epsilon^{klmn} (Z_{{\cal A}} -  Z_{\overline {\cal A}} ) \, \psi_l \sigma_m \overline \psi_n \, \partial_k a 
- \text{e}^{Z/2} (g \, \overline \psi_a \overline \sigma^{ab} \overline \psi_b + \overline g \, \psi_a \sigma^{ab} \psi_b ) 
\end{split}
\end{equation}
where the scalar potential is 
\begin{equation}
V = \text{e}^Z \left( f \overline f -3 g \overline g \right). 
\end{equation}
In all the expressions above one has to first calculate all the derivatives, and then set $X=0$ and ${\cal A} = \overline {\cal A} = a$. 

We end this section by discussing a simple application of this  constrained superfield to cosmology. 
Mimicking \cite{Kallosh:2014via,Dall'Agata:2014oka}, we choose K\"ahler potential and superpotential of the form
\begin{equation}
\begin{split}
K =& X \overline X - \frac14 ({\cal A} - \overline {\cal A})^2,
\\
W =& g({\cal A}) + X f({\cal A}),
\end{split}
\end{equation}
and
\begin{equation}
\overline{ f(z)} = f (\overline z) \ , \ \overline{ g(z)} = g (\overline z) .
\end{equation}
The complete theory in the $G_{\alpha}=0$ gauge is described by
\begin{equation}
\begin{split}
e^{-1} {\cal L} = & -\frac12 R
+ \frac{1}{2} \epsilon^{klmn} (\overline \psi_k \overline \sigma_l {\cal D}_m \psi_n - \psi_k \sigma_l {\cal D}_m \psi_n)
\\[2mm]
&
- \frac12 \partial^m a \, \partial_m a  - g(a) ( \overline \psi_a \overline \sigma^{ab} \overline \psi_b + \psi_a \sigma^{ab} \psi_b ) -V(a)
\end{split}
\end{equation}
and the scalar potential is given by
\begin{equation}
V = f(a)^2 - 3 g(a)^2.
\end{equation}
We will have an inflationary scenario with a susy breaking Minkowski vacuum if the following conditions hold at the critical point $a_0$
\begin{equation}
	V'(a_0) = 0 , 	\qquad f(a_0)^2 = 3 g(a_0)^2 \neq 0 
\end{equation}
and the functions $f$ and $g$ have been appropriately chosen. 
The scalar potential can be arbitrarily fixed in terms of these two functions, though one should carefully choose them in order for the effective theory to remain valid in the large range of scales touched during inflation and its exit period.
The supersymmetry breaking scale set by $F^x = f(a)$ fixes the cutoff for believing our effective theory, while the scale of quantum fluctuations is directly fixed by the scalar potential $V$.
The condition for perturbative unitarity during inflation is therefore fixed by requiring that $(F^x)^2 > V$, which is guaranteed for the model we are considering.



\section{A new constrained superfield} 
\label{sec:a_new_constrained_superfield}

In previous sections we discussed constrained superfields where all scalars in the models were projected out from the spectrum as well as models in which spin 1/2 fields where removed, while keeping the scalar fields in the same multiplet.
As previously noted, the last set of constraints removes also the auxiliary field of the same multiplet, producing Lagrangians for which the bosonic part cannot simply be obtained from the unconstrained version by setting to zero the constrained bosonic fields.
This means that, in order to extend such models to more general matter couplings, one has to work out from scratch all the couplings and the resulting physical consequences.

In this section we discuss an alternative constraint that is going to address this problem, by producing a multiplet where only the spin 1/2 field is removed, while the auxiliary field is still part of the game.
Also in this case the locally supersymmetric model is rather complicated, due to gravitational interactions, and we therefore discuss it first in global supersymmetry, so that we can better illustrate the properties of such constrained superfield.

We start again from a chiral multiplet ${\cal C}$, which in global supersymmetry has the expansion  
\begin{eqnarray}
{\cal C} = C(y) + \sqrt 2 \theta \chi(y) + \theta^2 F^{\cal C}(y),   
\end{eqnarray} 
where
\begin{eqnarray}
y_m = x_m + i \theta \sigma_m \overline \theta . 
\end{eqnarray}
We are interested in removing the fermionic component from the spectrum in a covariant way.
This requires to deal with the derived superfield
\begin{eqnarray}
\label{DC}
D_\alpha {\cal C} = \sqrt 2 \chi_\alpha(x) + \frac12 \theta_\alpha F^{\cal C}(x) 
+ 2 i \overline \theta^{\dot \alpha}  \sigma^m_{\alpha \dot \alpha} \partial_m C(x) + \cdots  . 
\end{eqnarray} 
Assuming a coupling to the goldstino multiplet $X$, so that $\chi$ can be expressed as a nontrivial function of the goldstino whenever the supersymmetry breaking parameter $F^x \neq 0$, we can extract the proper constraint by multiplying $D_{\alpha} {\cal C}$ with 
\begin{eqnarray}
\label{XbarX}
X \overline X = \theta^2 \overline \theta^2 F \overline F + {\cal O}(G, \overline G) . 
\end{eqnarray}
Only the highest component of $X \overline X$ contains a contribution that is independent from the Goldstino and therefore we can use it to remove $\chi$ from the spectrum. 
It is therefore natural to guess that a proper constraint would be
\begin{eqnarray}
\label{XDXC}
X \overline X D_\alpha {\cal C} = 0 . 
\end{eqnarray} 
However, we would like to impose a chiral constraint equivalent to this one, to be on the same footing as the other constraints presented in this work.
Using the nilpotency of the goldstino supefield $X$, which satisfies $X^2 = 0$, we can quickly see how to do this.
We can first build a composite vector field
\begin{equation}\label{compositeV}
	V = \overline{X} {\cal C} + X \overline{\cal C}, 
\end{equation}
which is also nilpotent $V^3 = 0$.
We then see that its field strength, defined by
\begin{equation}
	\widetilde{\cal W}_{\alpha} \equiv -\frac14 \overline{D}^2 D_{\alpha} V,
\end{equation}
is chiral.
A covariant chiral constraint is then
\begin{equation}
\label{XWT}
	X \widetilde{\cal W}_{\alpha} = -\frac14 \overline{D}^2 \left(X \overline X D_{\alpha} {\cal C}\right) = 0.
\end{equation}
From the point of view of $\widetilde{\cal W}_{\alpha}$ we have to solve the same equations as in section \ref{sub:models_with_a_single_u_1_vector}, which we know can be consistently solved by removing the fermion in $\widetilde{\cal W}_{\alpha}$ from the spectrum.
In this case this implies that the composite quantity $\widetilde{\cal W}_{\alpha}|_{\theta=0}$ is projected out, which in turn implies removing $\chi_{\alpha}$, given that
\begin{eqnarray}
\widetilde{\cal W}_{\alpha}|_{\theta=0}=  \sqrt 2 \overline F \chi_\alpha + \cdots .
\end{eqnarray}
Note that if we multiply equation \eqref{XWT} by $\overline X$  and use its nilpotency  we are left with 
\begin{equation}
\label{XWT2}
X \overline X \, \overline D^2 \overline X D_\alpha {\cal C} = 0.    
\end{equation}
Note also that $\overline D^2 \overline X$ is a  chiral superfield whose lowest component is never vanishing if supersymmetry is broken and therefore we can generically define its inverse. 
This means that \eqref{XWT2} implies \eqref{XDXC} and therefore the two versions of the constraint on ${\cal C}$ are equivalent. 

The coupling to supergravity does not change the argument, though it complicates the solution.
We start with the chiral superfield
\begin{eqnarray}
{\cal C} = C + \sqrt 2 \Theta \chi + \Theta^2 F^{\cal C},  
\end{eqnarray}
construct the composite vector superfield (\ref{compositeV}), its covariant field-strength
\begin{equation}
	\widetilde{\cal W}_{\alpha} = -\frac14  (\overline {\cal D}^2 - 8 {\cal R}) {\cal D}_\alpha V
\end{equation}
and impose the chiral constraint
\begin{equation}
	X \widetilde{\cal W}_{\alpha} =  -\frac14 (\overline {\cal D}^2 - 8 {\cal R}) \left(X \overline{X} {\cal D}_\alpha {\cal C}\right) = 0,
\end{equation}
which, also in this case, is equivalent to 
\begin{equation}
X \overline X {\cal D}_\alpha {\cal C} = 0 .
\end{equation}
We can then proceed and solve the constraint by taking its highest component by 
\begin{equation}\label{4Dnew} 
	{\cal D}^2 \overline {\cal D}^2  \left(X \overline{X} {\cal D}_\alpha {\cal C}\right)| = 0.
\end{equation}
The other components with less  derivatives are simply consistency conditions.
Note that the first term is just the highest $\theta$, $\overline{\theta}$ component in $X \overline{X} {\cal D}_{\alpha} C$, as expected.
From (\ref{4Dnew}) we find an equation of the form
\begin{eqnarray}
\chi_\alpha = B_\alpha  
+ A_{\alpha}^{\ \beta} \chi_\beta 
+\Gamma_\alpha^{\ m \beta} {\cal D}_m \chi_\beta 
+ \Delta_{\alpha}^{\ mn\beta} {\cal D}_m {\cal D}_n \chi_\beta . 
\end{eqnarray}
The reader interested in the complete expressions can find them  in the appendix, but for now it suffices to give the lowest order in Goldstino  (without derivatives hitting it) for the above coefficients. 
We have 
\begin{eqnarray}
 B_\alpha &=& {\cal O}(G) + {\cal O}(\overline G) ,
\\[2mm] 
A_{\alpha}^{\ \beta}   & =&  {\cal O}(G) + {\cal O}(\overline G)  ,
\\[2mm]
\Gamma_\alpha^{\ m \beta} & =&  {\cal O}(G \overline G)  ,
\\[2mm]
\Delta_{\alpha}^{\ mn\beta}& =&  {\cal O}(G^2 \overline G^2)  .   
\end{eqnarray}
By solving iteratively the resulting equation, we find   
\begin{eqnarray}
\chi_\alpha &= & B_\alpha 
+ A^{\ \beta}_{\alpha} B_\beta 
+ A_{\alpha}^{\ \beta} A_{\beta}^{\ \rho} 
\Big{[} 
B_\rho 
+ A_{\rho}^{\ \gamma} B_\gamma 
+\Gamma_{\rho}^{\ r \gamma} {\cal D}_{r} B_\gamma
 \Big{]}    
\\[2mm]
& 
+&
\Gamma_\alpha^{\ m \beta} {\cal D}_m B_\beta 
+ A_{\alpha}^{\ \beta} 
\Gamma_{\beta}^{\ k \rho} 
{\cal D}_k  
\Big{[} 
B_\rho 
+ A_{\rho}^{\ \gamma} B_\gamma 
+\Gamma_{\rho}^{\ r \gamma} {\cal D}_{r} B_\gamma
 \Big{]}    
\\[2mm]
&+&\Gamma_\alpha^{\ m \beta} {\cal D}_m 
\left( 
A_\beta^{\ \rho} 
\Big{[} 
B_\rho 
+ A_{\rho}^{\ \gamma} B_\gamma 
+\Gamma_{\rho}^{\ r \gamma} {\cal D}_{r} B_\gamma
 \Big{]} 
\right)
\\[2mm]
& +&\Gamma_\alpha^{\ m \beta} {\cal D}_m 
\left( 
\Gamma_\beta^{\ k \rho} {\cal D}_k 
\Big{[} 
B_\rho 
+ A_{\rho}^{\ \gamma} B_\gamma 
+\Gamma_{\rho}^{\ r \gamma} {\cal D}_{r} B_\gamma
 \Big{]}    
\right)
\\[2mm]
& +& \Delta_{\alpha}^{\ mn \beta} {\cal D}_m {\cal D}_n 
\Big{[} 
B_\beta  
+ A_{\beta}^{\ \rho} B_\rho 
+\Gamma_{\beta}^{\ k \rho} {\cal D}_{k} B_\rho
 \Big{]}    . 
\end{eqnarray}
This completes the solution and fixes the fermion $\chi_\alpha$ as a function of the other degrees of freedom of the theory. 

Removing only the fermion from the ${\cal C}$ multiplet may be puzzling, because the auxiliary field has not been removed and might get a non-vanishing vev. 
One would therefore expect the fermion associated to this auxiliary field to become a Goldstone mode. 
While this could be taken as a sign of an inconsistency, one can see that the first terms of the expression for the removed fermion are proportional to the $G_{\alpha}$ field
\begin{eqnarray}
\chi_\alpha = \frac{F^C}{F} G_\alpha + \frac{1}{F} \partial_{\alpha \dot \beta} C \, \overline G^{\dot \beta} + \cdots  
\end{eqnarray}
and, if $\langle F^C \rangle \ne 0$, the fermion $\chi$ contributes to the Goldstino, which is still aligned with $G_{\alpha}$:
\begin{eqnarray}
\chi_\alpha = \frac{\langle F^C \rangle}{\langle F \rangle} G_\alpha + \cdots .
\end{eqnarray}

We are once again interested in the component Lagrangian for the coupling of $X$ and ${\cal C}$ to supergravity.
In the $G=0$ gauge, 
\begin{eqnarray}\label{Cgauged}
{\cal C}|_{G=0} = C + \Theta^2 F^{{\cal C}} 
\end{eqnarray}
and all spin-1/2 fields vanish, so that the only fermion remaining is the gravitino. 
To illustrate the properties of the new superfield, we construct a simple model with K\"ahler potential and superpotential given by
\begin{eqnarray}
\begin{split}
K= &  X \overline X + Z({\cal C}, \overline {\cal C}) +  X \overline X Q({\cal C}, \overline {\cal C}),
\\[2mm]
W = & g({\cal C}) + X f({\cal C}) . 
\end{split}
\end{eqnarray}
The component form is quite simple in the $G=0$ gauge:
\begin{eqnarray} 
\begin{split}
e^{-1} {\cal L} = & -\frac12 R 
+ \frac{1}{2} \epsilon^{klmn} (\overline \psi_k \overline \sigma_l {\cal D}_m \psi_n - \psi_k \sigma_l {\cal D}_m \psi_n) 
- Z_{C \overline C} \partial C \partial \overline C 
\\[2mm]
& + \frac14 \epsilon^{klmn} Z_{{\cal C}} \, \psi_l \sigma_m \overline \psi_n \, \partial_k C
	 + \frac14 \epsilon^{klmn} Z_{\overline{\cal C}} \, \psi_l \sigma_m \overline \psi_n \, \partial_k \overline{C} \\[2mm]
& 
- \text{e}^{Z/2} \left( g(C) \overline \psi_a \overline \sigma^{ab} \overline \psi_b 
+ \overline g(\overline C)  \psi_a \sigma^{ab} \psi_b \right) 
- V ,
\end{split}
\end{eqnarray}
where the scalar potential has the form 
\begin{eqnarray}
V = e^K \left(|DW|^2 - 3W^2\right)_{x=0} = \text{e}^{Z} \left( \frac{|f|^2}{1 + Q} + \frac{|g_C + g Z_C|^2}{Z_{C \overline C }}   -  3 |g|^2  \right) . 
\end{eqnarray}

Using this setup one may easily obtain the complete component expressions for the models presented in \cite{Dall'Agata:2014oka}, while removing the fermion matter field in the inflationary sector.
To reproduce these models we set 
\begin{eqnarray}
\begin{split}
K =& X \overline X - \frac14 ({\cal C} - \overline {\cal C})^2 ,
\\
W =&  f( {\cal C}) \left( 1 + \sqrt{3} X \right),
\end{split}
\end{eqnarray}
where 
\begin{eqnarray}
\overline{f(z)} = f(\overline z) \ , \  f'(0) =0 \ , \ f(0) \ne 0 . 
\end{eqnarray}
The Lagrangian in component form is then
\begin{eqnarray} \label{lagrC}
\begin{split}
e^{-1} {\cal L} = & -\frac12 R 
+ \frac{1}{2} \epsilon^{klmn} (\overline \psi_k \overline \sigma_l {\cal D}_m \psi_n - \psi_k \sigma_l {\cal D}_m \psi_n) 
\\[2mm]
&
- \frac12 \partial^m C \, \partial_m \overline C 
-2 \text{e}^{-(C-\overline C)^2/4} \Big{|} f_C - \frac12 (C-\overline C) f(C) \Big{|}^2 
\\[2mm]
& -\frac18 (C-\overline C)\, \epsilon^{klmn} \, \psi_l \sigma_m \overline \psi_n \, \left(\partial_k C- \partial_k \overline{C} \right)\\[2mm]
&
- \text{e}^{- (C - \overline C)^2 /8} ( f(C) \overline \psi_a \overline \sigma^{ab} \overline \psi_b 
+ f(\overline C)  \psi_a \sigma^{ab} \psi_b ) . 
\end{split}
\end{eqnarray} 
Using this expression one can study the details of various models and further understand their cosmological properties. 
Note that the existence of a second scalar in \eqref{Cgauged} is in various scenarios welcome, as for example in the study related to the initial conditions problem as was done in \cite{Dalianis:2015fpa}. 
For single field inflation, the scalar in the imaginary component of $C$ has to be strongly stabilized (as explained in \cite{Kallosh:2014via,Dall'Agata:2014oka}) and we have 
\begin{eqnarray} 
\begin{split}
e^{-1} {\cal L} = & -\frac12 R 
+ \frac{1}{2} \epsilon^{klmn} (\overline \psi_k \overline \sigma_l {\cal D}_m \psi_n - \psi_k \sigma_l {\cal D}_m \psi_n) 
\\[2mm]
&
- \frac12 \partial^m c \, \partial_m c 
-2 |f'(c)|^2 -  f(c) \left(\overline \psi_a \overline \sigma^{ab} \overline \psi_b +   \psi_a \sigma^{ab} \psi_b \right) . 
\end{split}
\end{eqnarray} 


\section{Conclusions} 
\label{sec:conclusions}

In this note we considered minimal couplings of constrained superfields to supergravity.
We discussed the general solution of all known constraints where matter chiral multiplets transform in a non linear representation of supersymmetry in addition to the goldstino multiplet.
We have shown that supergravity generically modifies the solutions obtained previously in global supersymmetry, while reproducing the correct result in the rigid limit.

Having in mind applications of these models to particle physics or cosmology, we also provided their actions in components form for a specific gauge choice for the supersymmetry parameter, so that the fermion of the goldstino multiplet is set to zero: $G_{\alpha} = 0$.

Another interesting aspect is that models with constrained fermions have bosonic actions that cannot be simply obtained by setting to zero some of the terms in the corresponding unconstrained action.
For this reason we introduced a new constraint, which removes only the fermion in a chiral superfield ${\cal C}$, keeping its auxiliary field.
This is especially interesting, if we  think for instance of the Higgs multiplets in a low energy  non-linear supersymmetric theory, where the Higgsinos are removed from the spectrum. 
The fact that the auxiliary fields of the Higgs multiplets should not be removed from the spectrum a priori, can be traced to the supersymmetric Yukawa couplings and the $\mu$-term. 
These give contribution to the sfermion scalar potential and the Higgs mass when the Higgs auxiliary fields are eventually integrated out. 
Therefore, when using our constraint for the Higgs superfields in the low energy theory, one may still make use of the standard formulas of the MSSM for the scalar potential. 

There is clearly also plenty of room for applications of our results to cosmological models, where the scalar fields surviving the projections following from our constraints could be identified with the inflaton.
While we presented a couple of very simple models, for very specific choices of the K\"ahler potential and superpotential, one could try and see what happens for more general couplings. 
Of course, one should always be very careful with the validity of the resulting scalar potential at very high energy scales, because quantum corrections may destabilize the whole construction, giving contributions that go beyond the supersymmetry breaking scale, which sets the ultraviolet cutoff scale for our non-linear realizations.


\bigskip

\acknowledgments
We thank E.~Dudas, S.~Ferrara, A.~Kehagias, L.~Martucci and F.~Zwirner for discussions. This work was supported in part by the MIUR grant RBFR10QS5J (STaFI) and by the Padova University Project CPDA119349.

\appendix

\appendix

\section{The complete component expressions for the new constraint}

Here we present the complete expressions for the functions appearing in section \ref{sec:a_new_constrained_superfield}. 
For the Goldstino dependent coefficient $A_{\alpha}^{\ \gamma}$ we have   
\begin{align}
- 16 \sqrt{2} & F \overline F A_{\alpha}^{\ \gamma} = 
\\
\nonumber
& \sqrt 2 \delta_{\alpha}^{\gamma}   
[({\cal D}^\beta \overline {\cal D}^2 {\cal D}_\beta X|) ( \overline X |) 
+ 2 ({\cal D}^\beta \overline {\cal D}^{\dot \beta} {\cal D}_\beta X|) (\overline {\cal D}_{\dot \beta} \overline X|) 
+2 (\overline {\cal D}^{\dot \beta} {\cal D}_\beta X|)({\cal D}^\beta \overline {\cal D}_{\dot \beta} \overline X|) ] 
\\
\nonumber
& + 2 ({\cal D}^\beta \overline {\cal D}^{\dot \beta} {\cal D}_\beta X|) ( \overline X |) (\sqrt 2  i  \sigma^a_{\alpha \dot \beta} 
\psi_a^{\ \gamma} )  
+ 2 [ (\overline {\cal D}^{\dot \beta} {\cal D}_\beta X|) (\overline X|) ]  
( \sqrt 2 \overline \psi_{\alpha \dot \beta}^{\ \ \ \dot \rho} \epsilon^{\beta \rho} \psi_{\rho \dot \rho}^{\ \ \gamma} 
+ 2 \sqrt 2 i \, T^{\beta\ \ \gamma}_{\ \alpha \dot \beta} |   
)
\\
\nonumber
& 
- \sqrt 2  ({\cal D}_\beta X|) ({\cal D}^\beta \overline {\cal D}^2 \overline X|)   \delta_{\alpha}^{\gamma}   
- 2 ({\cal D}^2 X|) (\overline{\cal D}^{\dot \beta} \overline X |) (\sqrt 2 i \psi_{\alpha \dot \beta}^{\ \ \gamma} ) 
+ 2 ({\cal D}_\beta X|) ({\cal D}^\beta \overline {\cal D}^{\dot \beta} \overline X| ) (\sqrt 2 i \psi_{\alpha \dot \beta}^{\ \ \gamma} ) 
\\
\nonumber
& 
- 2 ({\cal D}_{\beta} X|)(\overline {\cal D}^{\dot \beta} \overline X|) 
(\sqrt 2 \overline \psi_{\alpha \dot \beta}^{\ \ \ \dot \rho} \epsilon^{\beta \rho} \psi_{\rho \dot \rho}^{\ \ \gamma} 
+ 2 i \sqrt 2 \, T^{\beta\ \ \gamma}_{\ \alpha \dot \beta} |)
+ ({\cal D}^2 X|)( \overline X|) \left( - \frac{4 \sqrt 2}{3} M \delta_{\alpha}^{\gamma} \right) 
\\
\nonumber
& 
- ({\cal D}_\beta X|)(\overline X|) 8 \sqrt 2 ({\cal D}^\beta {\cal R}|) \delta_{\alpha}^{\gamma} 
+ (X|) (\overline {\cal D}^2 \overline X|)  \frac{2 \sqrt 2}{3} \overline M \delta_{\alpha}^{\gamma} 
\\
\nonumber
& - 2 ({\cal D}^\beta X|)(\overline {\cal D}^{\dot \beta} \overline X| ) 
\left( - \sqrt 2 \overline \psi_{\alpha \dot \beta}^{\ \ \ \dot \alpha} \psi_{\beta \dot \alpha}^{\ \ \ \gamma} 
- 2 i \sqrt 2 T_{\beta \alpha \dot \beta}^{\ \ \ \ \gamma} | 
+ \sqrt 2 \overline \psi_{\beta \dot \beta}^{\ \ \ \dot \rho}   \psi_{\alpha \dot \rho}^{\ \ \ \gamma} 
- \sqrt 2 {\cal R}_{\dot \beta \beta \ \alpha}^{\ \ \ \gamma}|
\right) 
\\
\nonumber
& 
-2 (X|) ({\cal D}^\beta \overline {\cal D}^{\dot \beta} \overline X |)   
\left( - \sqrt 2 \overline \psi_{\alpha \dot \beta}^{\ \ \ \dot \alpha} \psi_{\beta \dot \alpha}^{\ \ \ \gamma} 
- 2 i \sqrt 2 T_{\beta \alpha \dot \beta}^{\ \ \ \ \gamma} | 
+ \sqrt 2 \overline \psi_{\beta \dot \beta}^{\ \ \ \dot \rho}   \psi_{\alpha \dot \rho}^{\ \ \ \gamma} 
- \sqrt 2 {\cal R}_{\dot \beta \beta \ \alpha}^{\ \ \ \gamma}|
\right) 
\\
\nonumber
& 
+ 2 (X|)(\overline {\cal D}^{\dot \beta} \overline X|) 
\left( 2i \sigma^c_{\alpha \dot \beta} 
\Big{[} \frac{2 \sqrt 2}{3} \overline M \psi_{c}^{\ \gamma} - \sqrt 2 i \overline \psi_c^{\ \dot \omega}  \epsilon^{\beta \rho} 
\frac{i}{2} \overline \psi^{\ \ \ \dot \rho}_{\rho \dot \omega}  \psi_{\beta \dot \rho}^{\ \ \ \gamma} 
- \frac{\sqrt 2}{3} \overline \psi_{c \dot \omega} b^{\gamma \dot \omega}  
\right. 
\\
\nonumber
& 
\ \ \ \ \ \ \ \ \ \ \ \ \ \ \ \ 
\ \ \ \ \ \ \ \ \ \ \ \ \ \ \ \  
- \frac{1}{\sqrt 2} \epsilon^{\beta \rho} T_{\rho c}^{\ \ \dot \delta}| \psi_{\beta \dot \delta}^{\ \ \ \gamma}  
\Big{]} 
- 2 i \sqrt 2 ({\cal D}^{\beta} T^{\ \  \ \  \gamma}_{\beta \alpha \dot \beta} |) 
\\
\nonumber
& 
\ \ \ \ \ \ \ \ \ \ \ \ 
\ \ \ \  \ \ \ \ \ \ \ \ 
- 2 i \sigma^c_{\beta \dot \beta} \epsilon^{\beta \rho} 
\Big{[}   \frac{ \sqrt 2}{3} \epsilon_{\rho \alpha} \overline M \psi_{c}^{\ \gamma} 
- \sqrt 2 i \overline \psi_c^{\ \dot \beta}  
\frac{i}{2} \overline \psi^{\ \ \ \dot \rho}_{\rho \dot \beta}  \psi_{\alpha \dot \rho}^{\ \ \ \gamma} 
- \frac{1}{3 \sqrt 2} \epsilon_{\rho \alpha} \overline \psi_{c \dot \beta} b^{\gamma \dot \beta}  
\\
\nonumber
& 
\ \ \ \ \ \ \ \ \ \ \ \ \ \ \ \ 
\ \ \ \ \ \ \ \ \ \ \ \ \ \ \ \  
\ \ \ \ \ \ \ \ \ 
\left. 
- \frac{1}{\sqrt 2} T_{\rho c}^{\ \ \dot \delta}| \psi_{\alpha \dot \delta}^{\ \ \ \gamma}  
\Big{]} 
- \sqrt 2 ({\cal D}^{\beta} {\cal R}^{\ \  \ \gamma}_{\dot \beta \beta \ \alpha} |) 
\right) 
\\
\nonumber
& 
+ ({\cal D}^{\beta} X|)(\overline X|) \frac12 \epsilon_{\beta \alpha}  \Big{[} 
- \frac{16}{\sqrt 2} e_{a}^{m} {\cal D}_m \psi^{a \gamma} + 
\frac{16 i \sqrt 2}{3} b^a \psi_a^{\ \gamma}
- 8 \sqrt 2 \psi^{m \rho}\frac{i}{2} \overline \psi_{m}^{\ \dot \omega} \psi_{\rho \dot \omega}^{\ \ \  \gamma}   
\\
\nonumber
& 
\ \ \ \ \ \ \ \ \ \ \ \ 
\ \ \ \  \ \ \ \ \ \ \ \  
\ \ \ \ 
- \frac{8 \sqrt 2}{3} \sigma_{\ \ \ \rho}^{nm \ \omega} \psi_{mn \omega} \epsilon^{\rho \gamma} 
- \frac{8 i \sqrt 2}{3} b^a \psi_a^{\ \gamma} 
- \frac{4 i \sqrt 2}{3} b_{\rho \dot \rho} \overline \sigma^{a \dot \rho \epsilon} \psi_{a \epsilon} 
\epsilon^{\rho \gamma} 
\Big{]} . 
\end{align}
For the coefficient $B_\alpha$ we have    
\begin{align}
- 16 \sqrt 2 & F \overline F  B_\alpha = 
\\
\nonumber
& 2 (\overline {\cal D}^2 {\cal D}_{\alpha} X| )(\overline X|)  F^C 
+ 4 (\overline {\cal D}^{\dot \beta} {\cal D}_{\alpha} X| )(\overline {\cal D}_{\dot \beta} \overline X|)   F^C 
+ 2 ({\cal D}_\alpha X|) (\overline {\cal D}^2 \overline X|)  F^C 
\\
\nonumber
& + 2 ({\cal D}^\beta \overline {\cal D}^{\dot \beta} {\cal D}_\beta X|)(\overline X|) (- 2 i \partial_{\alpha \dot \beta} C)  
+ 4 i (\overline {\cal D}^{\dot \beta} {\cal D}_\beta X| ) (\overline X|) 
 \left( \psi_{\alpha \dot \beta}^{\ \ \ \beta} F^C 
- \overline \psi_{\alpha \dot \beta}^{\ \ \ \dot \rho} \partial_{\ \dot \rho}^{\beta} C \right) 
\\
\nonumber
&
+ 2 ({\cal D}_\beta X|)({\cal D}^\beta \overline {\cal D}^{\dot \beta} \overline X|)  
(-2i \partial_{\alpha \dot \beta} C)   
-4 i ({\cal D}_\beta X|)( \overline {\cal D}^{\dot \beta} \overline X| )
\left( \psi_{\alpha \dot \beta}^{\ \ \ \beta} F^C 
- \overline \psi_{\alpha \dot \beta}^{\ \ \ \dot \rho} \partial_{\ \dot \rho}^{\beta} C \right) 
\\
\nonumber
& 
- ({\cal D}_\alpha X|)(\overline X|) \frac83  F^C M  
+  ({\cal D}^\beta X|)(\overline {\cal D}^2 \overline X|) (- 2 \epsilon_{\beta \alpha} F^C)  
+ (X|) ({\cal D}^\beta \overline {\cal D}^2 \overline X|) (- 2 \epsilon_{\beta \alpha} F^C ) 
\\
\nonumber
& 
+4 \left\{ ({\cal D}^\beta X|)(\overline {\cal D}^{\dot \beta} \overline X| ) +  (X| )({\cal D}^\beta \overline {\cal D}^{\dot \beta} \overline X|) \right\}
\Big{[} 
i \psi_{\alpha \dot \beta \beta} F^C 
- \overline \psi_{\alpha \dot \beta}^{\ \ \ \dot \alpha} \partial_{\beta \dot \alpha} C 
\\
\nonumber
& 
\ \ \ \ \ \ \ \ \ \ \ \
\ \ \ \ \ \ \ \ \ \ \ \ \ \   
\ \ \ \ \ \ \  \   \  \  \  \  
\ \ \ \ \ \ \ \ \ \ \ \ \ \ \ 
\ \ \ 
- i \psi_{\beta \dot \beta \alpha} F^C 
+ \overline \psi_{\beta \dot \beta}^{\ \ \ \dot \rho} \partial_{\alpha \dot \rho} C \Big{ ] } 
\\
\nonumber
& 
+ 2 (X| )(\overline {\cal D}^{\dot \beta} \overline X |) {\cal K}_{\alpha \dot \beta}  
+  ({\cal D}^\beta X|) ( \overline X|) \frac12 \epsilon_{\beta \alpha} {\cal P}     
+ 4 i ({\cal D}^2 X|)(\overline {\cal D}^{\dot \beta} \overline X| )  \partial_{\alpha \dot \beta} C 
\\
\nonumber
& 
- \frac12 (X|)(\overline X|) \Big{[} 16 (\overline {\cal D}_{\dot \beta} \overline {\cal R} |) \partial_{\alpha}^{\ \dot \beta} C 
- 4 i \sigma^m_{\alpha \dot \alpha} \epsilon^{ \dot \alpha \dot \rho} 
{\cal D}_m  ( - 4 i \psi_{\beta \dot \rho }^{\ \ \ \beta} F^C 
- 4 \overline \psi_{\beta \dot \rho \dot \omega} \partial^{\beta \dot \omega} C ) 
\\
\nonumber
& 
\ \ \ \ \ \ \ \ \ \ 
\ \ \ \ \ \ \ \ \ 
- 4 i \epsilon^{\dot \alpha \dot \beta} \psi_{\alpha \dot \alpha}^{\ \  \ \mu} {\cal K}_{\mu \dot \beta } 
- i \epsilon^{\dot \alpha \dot \rho} \overline \psi_{\alpha \dot \alpha \dot \rho} \,  {\cal P}  
\\
\nonumber
& 
\ \ \ \ \ \ \ \ \ \ 
\  \ \ \ \ \  \ \ \ \ \ 
+2 i \epsilon^{\dot \alpha \dot \rho} T_{\dot \rho \alpha \dot \alpha \dot \delta}| \epsilon^{\dot \delta \dot \beta} 
(-4 i  \psi_{\beta \dot \beta}^{\ \ \ \beta} F^C 
- 4  \overline \psi_{\beta \dot \beta \dot \rho} \partial^{\beta \dot \rho} C  )   
\\
\nonumber
&
\ \ \ \  \ \  \ \   \ 
\ \ \ \ \ \ \ \ \ \ \ \ 
- ({\cal R}_{\alpha \dot \alpha}^{\ \ \ \dot \beta \dot \alpha} | ) 
(-4 i  \psi_{\beta \dot \beta}^{\ \ \ \beta} F^C 
- 4  \overline \psi_{\beta \dot \beta \dot \rho} \partial^{\beta \dot \rho} C  )   \Big{ ] } . 
\end{align}
Here we have used the expressions 
\begin{eqnarray}
\begin{split}
{\cal K}_{\alpha \dot \beta} = & 2i \sigma^c_{\alpha \dot \beta} \Big{[} -4e_c^m {\cal D}_m F^C 
- \sqrt 2 i \overline \psi_{c}^{\dot \omega} \epsilon^{\beta \rho} \sigma_{\rho \dot \omega}^{m} 
(-\frac{1}{\sqrt 2} \psi_{m \beta} F^C - \frac{i}{\sqrt 2} \overline \psi_{m}^{\ \ \dot \rho} \partial_{\beta \dot \rho} C ) 
\\
&  \  \  \  \  \ \ \ \  \  \ \  \   \    
- 2 \epsilon^{\beta \rho} T_{\rho c \beta}| F^C  
+ \epsilon^{\beta \rho} T_{\rho c}^{\ \ \dot \delta} | 
\partial_{\beta \dot \delta}  C  \Big{]} 
+ 4 i  \epsilon^{\beta \rho} T_{\beta \alpha \dot \beta \rho}| F^C  
\\
& - 2 i \sigma^c_{\beta \dot \beta} \epsilon^{\beta \rho} 
\Big{[} - 2 \epsilon_{\rho \alpha} e_{c}^{m} \partial_m F^C 
- \sqrt 2 i \overline \psi_{c}^{\dot \beta } \sigma^m_{\rho \dot \beta} 
( -\frac{1}{\sqrt 2} \psi_{m \alpha} F^C - \frac{i}{\sqrt 2} \overline \psi_{m}^{\ \dot \rho} \partial_{\alpha \dot \rho} C) 
\\
&
\ \ \ \ \ \ \  \ \ \ \  
\ \  \ \  \ \ \
-2 T_{\rho c \alpha}| F^C + T_{\rho c}^{\ \ \dot \delta} | \partial_{\alpha \dot \delta} C \Big{]}  
+ 2 \epsilon^{\beta \delta} F^C {\cal R}_{\dot \beta \beta \delta \alpha}|
\end{split}
\end{eqnarray}
and
\begin{eqnarray}
{\cal P} = 16 e_{a}^{m} {\cal D}_m (e^{an} \partial_n C  ) 
- \frac{32 i }{3}b^n \partial_n C 
+8 \psi^{m \rho} ( \psi_{m \rho} F^C +i \overline \psi_{m}^{\ \dot \omega} \partial_{\rho \dot \omega} C  ) 
+ \frac{32}{3} M F^C . 
\end{eqnarray}
For $\Gamma_{\alpha}^{\ p \gamma}$ we have  
\begin{align} 
- 16 \sqrt 2 & F \overline F \,  \Gamma_{\alpha}^{\ p \gamma} = 
\\
\nonumber
&  2 (\overline {\cal D}^{\dot \beta} {\cal D}_\beta X |)(\overline X|) (-2 i \sqrt 2 \sigma^p_{\alpha \dot \beta} \epsilon^{\beta \gamma}) 
- 2 ({\cal D}_\beta X|)(\overline {\cal D}^{\dot \beta} \overline X|) (- 2i \sqrt 2 \sigma^p_{\alpha \dot \beta} \epsilon^{\beta \gamma} )  
\\
\nonumber
& -2 [({\cal D}^\beta X|)(\overline {\cal D}^{\dot \beta} \overline X|) + (X|)({\cal D}^\beta \overline {\cal D}^{\dot \beta} \overline X|) ]
(2 i \sqrt 2 \sigma^p_{\alpha \dot \beta} \delta_{\beta}^{\gamma} 
-2i \sqrt 2 \sigma^p_{\beta \dot \beta} \delta_{\alpha}^{\gamma}) 
\\
\nonumber
& 
+ 2 (X|) (\overline {\cal D}^{\dot \beta} \overline X) 
(2 \sqrt 2 \overline \psi_{\alpha \dot \beta}^{\ \ \ \dot \omega} \sigma^p_{\rho \dot \omega} \epsilon^{\gamma \rho} 
- 2 \sqrt 2 \epsilon^{\beta \rho} \overline \psi_{\beta \dot \beta}^{\ \ \ \dot \beta} \sigma^p_{\rho \dot \beta} \delta_{\alpha}^{\gamma}) 
\\
\nonumber
&+  ({\cal D}^\beta X|) (\overline X| ) \frac12 \epsilon_{\beta \alpha} (- 16 \sqrt 2 \psi^{p \gamma} )  
\\
\nonumber
& - \frac{1}{2} (X|)(\overline X|) \Big{[} 16 \sqrt 2 \sigma^m_{\alpha \dot \alpha} \epsilon^{\dot \alpha \dot \rho} 
{\cal D}_m(e_{d}^{\ p}) \sigma^{d}_{\beta \dot \rho} \epsilon^{\beta \gamma} 
- i \overline \psi_{\alpha \dot \alpha}^{\ \ \ \dot \alpha} (- 16 \sqrt 2 \psi^{p \gamma} )   
\\
\nonumber
&
\ \ \ \ \ \ \ \ \ \ \ \ \  
\ \ \ \ \ \ 
- 4 i \sigma^e_{\alpha \dot \alpha} \epsilon^{\dot \alpha \dot \beta} \psi_e^{\mu}  
(2 \sqrt 2 \overline \psi_{\mu \dot \beta}^{\ \ \ \dot \omega} \sigma^p_{\rho \dot \omega} \epsilon^{\gamma \rho} 
- 2 \sqrt 2 \epsilon^{\beta \rho} \overline \psi_{\beta \dot \beta}^{\ \ \ \dot \beta} \sigma^p_{\rho \dot \beta} \delta_{\mu}^{\gamma}) 
\\
\nonumber
&
\ \ \ \ \ \ \ \ \ \ \ \ \ \ \ 
\ \ \ \ \ 
+ ( 2 i \epsilon^{\dot \alpha \dot \rho} T_{\dot \rho \alpha \dot \alpha \dot \delta}| \epsilon^{\dot \delta \dot \beta} 
- {\cal R}_{\alpha \dot \alpha}^{\ \ \ \dot \beta \dot \alpha}| ) (4 i \sqrt 2 \sigma^p_{\beta \dot \beta} \epsilon^{\beta \gamma}) 
\Big{]} . 
\end{align}
Finally for $\Delta_{\alpha}^{\ pq\gamma}$ we have found 
\begin{eqnarray}
- 16 \sqrt 2 F \overline F \Delta_{\alpha}^{\ pq\gamma} = - 8 \sqrt 2 (X|)(\overline X|) \sigma^p_{\alpha \dot \alpha} 
\epsilon^{\dot \alpha \dot \rho} 
\sigma^q_{\beta \dot \rho} \epsilon^{\beta \gamma} . 
\end{eqnarray}
The various Goldstino contributions can be extracted by the 
expressions for the Goldstino superfield in the second section,  
and the following component expressions 
\begin{align}
{\cal D}^\beta \overline {\cal D}^2 {\cal D}_\beta X| =& 8 \sqrt 2 ({\cal D}^\beta {\cal R}|) G_\beta + \frac{16}{3} M F  ,
\\
\overline {\cal D}^2 {\cal D}_\beta X| =& - \frac{4 \sqrt 2}{3} M G_\beta  ,
\\
{\cal D}^\beta \overline {\cal D}^{\dot \beta} {\cal D}_\beta X| =& 2 i \sqrt 2 {\widehat D}^{\dot \beta \beta} G_\beta 
- \frac{5 \sqrt 2}{3} b^{\dot \beta \rho} G_\rho, 
\\
{\cal D}^\beta \overline {\cal D}^2 \overline X | =& 4 i \sqrt 2 {\widehat D}^{\beta \dot \gamma} \overline G_{\dot \gamma} 
+ \frac{2 \sqrt 2}{3} b^{\beta \dot \gamma} \overline G_{\dot \gamma} , 
\\
{\cal D}_\beta \overline {\cal D}_{\dot \beta} \overline X| =& - 2 i {\widehat D}_{\beta \dot \beta} \left( \frac{\overline G^2}{2 \overline F} \right). 
\end{align}
Regarding the superspace curvatures and superspace torsion components which appear in these expressions, 
these are the standard ones for old-minimal supergarvity and can be found for example in reference \cite{Wess:1992cp}.

\end{document}